# Tunable electronic and magneto-optical properties of monolayer arsenene: from $GW_0$ approximation to large-scale tight-binding propagation simulations

Jin Yu[1,2], Mikhail I. Katsnelson[2] and Shengjun Yuan[3,1,2]

1 Beijing Computational Science Research Center, Beijing 100094, China

2 Theory of Condensed Matter, Radboud University, Nijmegen 6525AJ, The Netherlands

3 School of Physics and Technology, Wuhan University, Wuhan 430072, China

**Abstract**

Monolayers of group VA elements have attracted great attention with the rising of black phosphorus. Here, we derive a simple tight-binding model for monolayer grey arsenic, referred as arsenene (ML-As), based on the first-principles calculations within the partially self-consistent $GW_0$ approach. The resulting band structure derived from the six *p*-like orbitals coincides with the quasi-particle energy from $GW_0$ calculations with a high accuracy. In the presence of a perpendicular magnetic field, ML-As exhibits two sets of Landau levels linear with respect to the magnetic field and level index. Our numerical calculation of the optical conductivity reveals that the obtained optical gap is very close to the $GW_0$ value and can be effectively tuned by external magnetic field. Thus, our proposed TB model can be used for further large-scale simulations of the electronic, optical and transport properties of ML-As.

**Introduction**

In the past few years, two-dimensional (2D) materials have attracted great attention, from graphene to transition metal dichalcogenides, black phosphorus, layered boron, mixed metal carbides etc. [1-5] As a precursor of monolayers composed of group VA elements, black phosphorus is reported to have potential applications in optoelectronics devices and field effect transistors, owing to its appropriate band gap and excellent electronic properties including a high carrier mobility and on-off current ratio.[3] First-principles calculations on these group VA element monolayers have shown that their energy gap covers a wide range from 0.36 to 2.62 eV with the carrier mobility varying from ~10 to ~$10^4$ $cm^2V^{-1}s^{-1}$. [6-9] One of the group VA elements is arsenic, the binary compound of which, gallium arsenide (GaAs), has been widely known in semiconducting industry in the last century. [10,11] Meanwhile, monolayer arsenic, namely arsenene, is receiving continuous interests very recently. [12-16] When it is exploited from bulk gray arsenic into few layers, even into monolayer arsenene (ML-As), it undergoes a transition from metal to semiconductor. [17] The indirect energy gap can be tuned into direct energy gap under biaxial strain. [12,17,18] When the strain is large enough (around 11.7%), theoretically, monolayer arsenene will become nontrivial topological insulator with a sizable bulk gap up to several tenths of 1 eV, [19] which is much higher than traditional 2D materials like graphene,[20] silicone, and germanene by one or two orders of magnitude. [21,22] External electric field, which is used as an effective way to tune the electronic properties, is found to introduce topological phase transition in ML-As. [23] It is also reported that by passivation of halogen atoms and functional groups like $CH_3$ and OH, one can introduce quantum spin Hall effect in arsenene. [24-26] Moreover, strain-tunable magnetism is predicted in ML-As with either electron or hole doping. [27]

The physical properties of ML-As have been studied extensively by density functional theory (DFT) investigations, [28,29] which is applicable only for the systems with not too high number of nonequivalent atomic positions (up to $10^3$). On the contrary, the electronic, optical, transport and plasmonic properties of various 2D materials can be studied efficiently within the framework of tight-binding approximation without any symmetry restrictions up to $10^9$ atoms. [30]

There have been successfully fitted tight-binding models for group VA elements P and Sb, [31,32] however, an effective tight-binding model for ML-As is still missing. Considering the fact that the binding energy of the first bound exciton of ML-As is evaluated on the order of 0.7 eV when the electron-hole interaction is taken into account, [33] many-body effect is crucially important in altering the optical properties of this 2D material, so it is essential to build a tight-binding model (TB) for ML-As based on the quasi-particle energy beyond density functional.

In this paper, we present a suitable model Hamiltonian governing the low-energy band structure of ML-As and show that its electronic and magneto-optical properties can be tuned by a perpendicular magnetic field. Our first-principles calculations show that the low-energy electronic properties of ML-As is mainly contributed by the 3*p* orbitals of As atom. Therefore a set of hybrid *p*-like orbitals generated by maximally localized Wannier functions [34,35] are chosen as the basis to construct the simple TB model for ML-As. The resulting band structure and electron density of states are in good agreement with the one obtained from the interpolation of the results obtained from the $GW_0$ calculations. We further use the fitted TB model to study the electronic and optical properties of large-scale ML-As under the external magnetic field, in which the ML-As exhibits highly degenerate equidistant Landau levels and tunable magneto-optical conductivities.

**Computational Method**

To construct a reliable TB model for ML-As, we performed first-principles calculations to calibrate the effective Hamiltonian, which is carried out by the Vienna Ab-initio simulation package (VASP) code. [36,37] The vacuum distance between two adjacent ML-As was set to be at least 1.5 nm to avoid interlayer interaction. The generalized gradient approximation [38] was used in combination with the projected augmented-wave method [39]. The plane wave cutoff energy was set to 260.9 eV. A 25*25*1 Monkhorst-Pack grid was used for the Brillouin zone sampling for both the relaxation and static calculations. All the atoms are fully relaxed until the force on each atom was less than 0.01 eV/Å. Wannier90 code [40] was used to construct the Wannier functions and TB parametrization of the DFT Hamiltonian. The obtained hopping parameters are further discard and re-optimized through a least-square fitting of the band structure. The electronic density of sates (Landau level spectrum) and the magneto-optical conductivity under external magnetic field was calculated using the tight-binding propagation method (TBPM), which is based on the numerical solution of the time-dependent Schrödinger equation without the diagonalization of the Hamiltonian matrix, and very efficient in the calculations of the electronic, transport an optical properties of large quantum systems with more than millions of atoms. [30,41,42]

**Results and discussion**

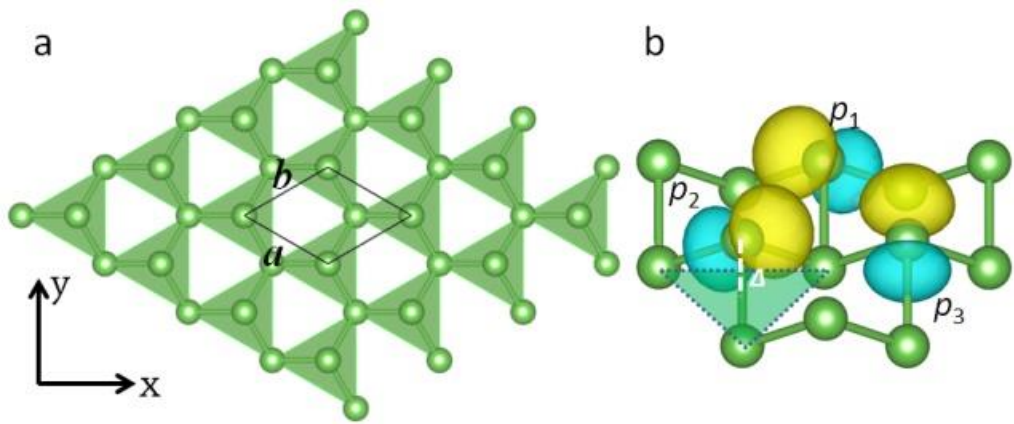

**Figure 1** Atomic structure of ML-As. (a) Top view of ML-As with the unit cell indicated by the black rhombus. The green balls are arsenic atoms. (b) Real-space distribution of the WFs at one sublattice, where three *p*-like orbitals are situated.

The optimized atomic structure of ML-As is shown in Figure 1, which has a hexagonal unit cell with the space group being $R\bar{3}m$. The basic unit block contains two sublattices occupied by one As atom per site. After structure optimization, the lattice constant of ML-As is calculated to ***a=b*=3.61 Å** with the out-of plane buckling of two sites being *Δ*=1.40 Å, which is in agreement with previous DFT calculations. [33] Standard DFT calculation shows that ML-As is a semiconductor with an indirect energy gap around 1.60 eV, and the valence band maximum (VBM) situates right at the high symmetric Γ point while the conduction band minimum (CBM) situates between Γ and M. The orbital decomposed band structure from DFT calculations shows that the first three valence and conduction bands are mainly composed of the 3*p* orbitals of As and that they are separated from the other states, which makes it possible for us to build an accurate description of effective Hamiltonian for ML-As. Thus, we will focus on the six bands in the low-energy regime. To get a much accurate description of the quasi-particle energy, we further performed $GW_0$ calculations on the electronic structure. A much larger energy gap is calculated to be 2.75 eV. Our parametrization procedure in this work is based on the formalism of maximally localized Wannier functions (MLWFs) [34,35] and six hybrid *p*-like orbitals are obtained as the basis for the TB model. In Figure 1b, we visualize the real-space distribution of the WFs at one site, where a combination of three equivalent *p*-like orbitals are situated around at the same site of As with a rotation symmetry of $2\pi/3$.

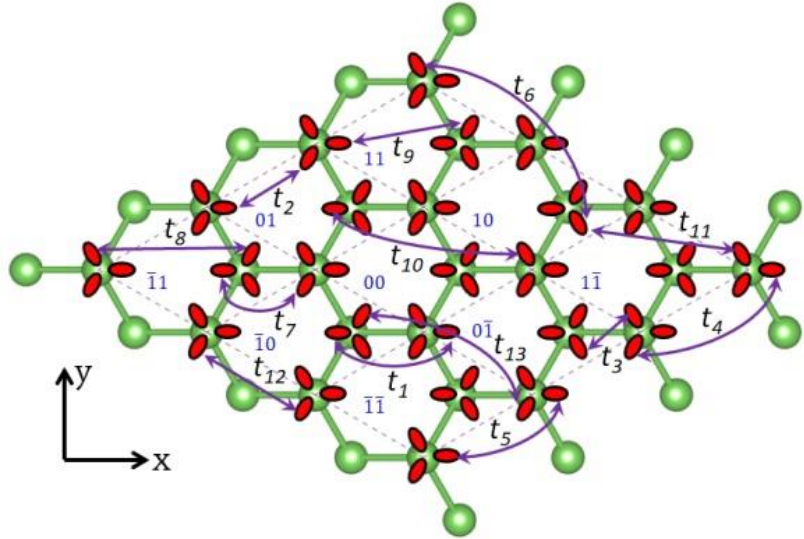

**Figure 2** Hopping diagram in the TB model of ML-As. The orbitals are represented by the negative part of the *p*-like orbitals as indicated by the red ellipses.

Our nonrelativistic TB model is given by an effective Hamiltonian of

$$H_0 = \sum_{mn} \sum_{ij} t_{ij}^{mn} a_{im}^\dagger a_{jn}, \quad (1)$$

where $t_{ij}^{mn}$ is the effective hopping parameter describing the interaction between $m$ and $n$ orbitals residing at atoms $i$ and $j$, respectively, and $a_{im}^\dagger$ ($a_{jn}$) is the creation (annihilation) operator of an electron at atom $i$ ($j$) and orbital $m$ ($n$). To make the TB model simple but accurate enough, we discard the hoppings with an interatomic distance larger than 5.69 Å and the hoppings with amplitudes $|t_i| <$ 0.07 eV are ignored as well. The residual hopping parameters are further re-optimized by minimize the following least square function

$$\delta(\{t_i\}) = \sum_{n,k} [E_{n,k}^{GW_0}(\{t_i\})^2 - E_{n,k}^{TB}(\{t_i\})^2] / |E_{n,k}^{GW_0}(\{t_i\})|, \quad (2)$$

where $\{t_i\}$ are the hoppings in equation (1) and $E_{n,k}^{GW_0}(\{t_i\})$ ( $E_{n,k}^{TB}(\{t_i\})$ ) corresponds to the eigenvalues of $GW_0$ (TB) Hamiltonian with $n$ and $k$ being the band index and momenta along the high-symmetry $K$ points in the first Brillouin zone, respectively. The final hoppings $\{t_i\}$ used in our calculations are schematically shown in Figure 2 and Table 1.

**Table 1** Hopping parameters with relevant hopping distance in the ML-As TB model.

| i | $t_i$(eV) | d(Å) | i | $t_i$(eV) | d(Å) | i | $t_i$(eV) | d(Å) |
|---|---|---|---|---|---|---|---|---|
| 1 | -3.52 | 2.51 | 6 | -0.26 | 5.69 | 11 | 0.10 | 5.69 |
| 2 | -0.76 | 3.61 | 7 | 0.12 | 2.51 | 12 | -0.09 | 3.61 |
| 3 | 0.72 | 2.51 | 8 | -0.15 | 4.40 | 13 | -0.07 | 5.69 |
| 4 | 0.27 | 3.61 | 9 | -0.10 | 4.40 | | | |
| 5 | 0.25 | 3.61 | 10 | 0.10 | 5.69 | | | |

In equation (1), we define the Hamiltonian with a 6*6 matrix for ML-As which can be explicitly solved to get the eigenvalues and eigenvectors. Considering the inversion symmetry, the Hamiltonian matrix can be simply written as

$$H(\mathbf{k}) = \begin{pmatrix} U(\mathbf{k}) & T(\mathbf{k}) \\ T^\dagger(\mathbf{k}) & U(\mathbf{k}_r) \end{pmatrix}, \quad (3)$$

where $U(\mathbf{k})$ and $T(\mathbf{k})$ are 3*3 matrices describing the intra- and inter-sublattice hoppings, respectively, the subscript $r$ in $U(\mathbf{k}_r)$ indicates the inversion operation in $U(\mathbf{k})$. As the three basic orbitals are rotationally symmetric like a regular triangle, the corresponding matrices in equation (3) are expanded as

$$U(\mathbf{k}) = \begin{pmatrix} A(\mathbf{k}) & B(\mathbf{k}) & B^*(\bar{\bar{\mathbf{k}}}) \\ B^*(\mathbf{k}) & A(\bar{\mathbf{k}}) & B(\bar{\mathbf{k}}) \\ B(\bar{\bar{\mathbf{k}}}) & B^*(\bar{\mathbf{k}}) & A(\bar{\bar{\mathbf{k}}}) \end{pmatrix}, \quad (4)$$

and

$$T(\mathbf{k}) = \begin{pmatrix} C(\mathbf{k}) & D(\mathbf{k}) & D(\bar{\bar{\mathbf{k}}}) \\ D(\mathbf{k}) & C(\bar{\mathbf{k}}) & D(\bar{\mathbf{k}}) \\ D(\bar{\bar{\mathbf{k}}}) & D(\bar{\mathbf{k}}) & C(\bar{\bar{\mathbf{k}}}) \end{pmatrix}, (5)$$

where $\bar{\mathbf{k}}$ and $\bar{\bar{\mathbf{k}}}$ are the $\mathbf{k}$ vector rotated by $2\pi/3$ and $-2\pi/3$, respectively. And the matrix elements in equation (4) and (5) read

$$A(\mathbf{k}) = 2t_{12}\cos(k_y a) + 2t_5 e^{\frac{\sqrt{3}}{2}ik_x a}\cos(\frac{1}{2}k_y a), (6)$$

$$B(\mathbf{k}) = t_2 e^{-i(\frac{1}{2}k_y a - \frac{\sqrt{3}}{2}k_x a)} + t_4 e^{i(\frac{1}{2}k_y a - \frac{\sqrt{3}}{2}k_x a)}, (7)$$

$$C(\mathbf{k}) = t_1 e^{-\frac{2\sqrt{3}}{3}ik_x a} + 2t_3 e^{\frac{\sqrt{3}}{6}ik_x a}\cos(\frac{1}{2}k_y a) + 2t_{14} e^{\frac{\sqrt{3}}{6}ik_x a}\cos(\frac{3}{2}k_y a) + 2t_6 e^{-\frac{5\sqrt{3}}{6}ik_x a}\cos(\frac{1}{2}k_y a) + 2t_{11} e^{\frac{2\sqrt{3}}{3}ik_x a}\cos(k_y a) + 2t_{13} e^{\frac{\sqrt{3}}{6}ik_x a}\cos(\frac{3}{2}k_y a), (8)$$

$$D(\mathbf{k}) = t_7 e^{-i\frac{\sqrt{3}}{3}k_x a} + t_7 e^{i(\frac{\sqrt{3}}{6}k_x a - \frac{1}{2}k_y a)} + t_8 e^{-i(\frac{\sqrt{3}}{3}k_x a + k_y a)} + t_9 e^{\frac{2\sqrt{3}}{3}ik_x a} + t_{10} e^{-i(\frac{5\sqrt{3}}{6}k_x a - \frac{1}{2}k_y a)}. (9)$$

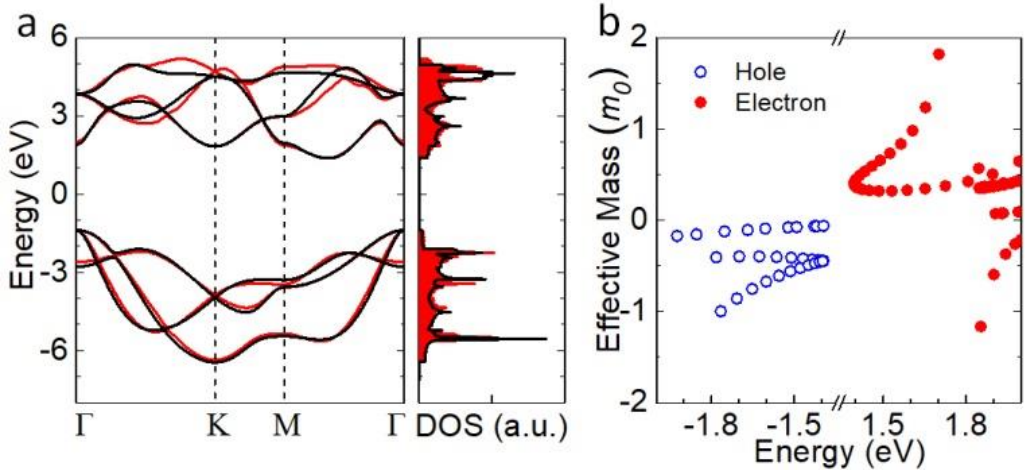

**Figure 3** Electronic properties of ML-As. (a) Band structure and DOS of ML-As. The red and black colors present the $GW_0$ and TB results, respectively. (b) Effective mass of carriers derived from the TB model in the unit of free electron mass $m_0$.

The EMs of the first CB (electron) and first two VBs (hole) are presented in red and blue, respectively.

The band structure and density of states (DOS) calculated from the effective Hamiltonian is shown in Figure 3a, where the quasi-particle energy within $GW_0$ approximation is plot as a reference. One can see a quite accurate match between the TB model and first-principles calculations. The six bands are almost overlapped with the 2$^{nd}$ and 3$^{rd}$ conduction bands slightly shifted away from the $GW_0$ bands. Particularly, the VBM is situated right at the Γ point and the CBM is located between the Γ and M point, rendering an indirect energy gap. It is also noted that the band structure matches quite well between -3 eV and 3 eV, resulting in a good agreement of the DOS.

**Table 2** Energy gaps and carrier effectives of ML-As. The indirect (ΓM) and direct (ΓΓ) energy gaps $E_g$ (in eV) and effective mass for hole and electron (in unit of the free electron mass $m_0$) at relevant high-symmetry $K$ points of the first Brillouin zone are listed. Γ1 and Γ2 represent the first and second band at the Γ point.

| Method | Energy gap (eV) | | Holes ($m_0$) | | Electrons ($m_0$) | | |
|---|---|---|---|---|---|---|---|
| | $E_g^{\Gamma M}$ | $E_g^{\Gamma \Gamma}$ | $m^{\Gamma 1}$ | $m^{\Gamma 2}$ | $m^{\Gamma}$ | $m^{\Gamma M}$ | $m^K$ |
| DFT-GW | 2.75 | 3.38 | 0.40 | 0.08 | 0.09 | 0.39 | 0.37 |
| TB | 2.79 | 3.30 | 0.44 | 0.06 | 0.07 | 0.41 | 0.35 |

The accuracy of the TB model for ML-As is quantitatively evaluated by analyzing the energy gap and carrier effective mass (EM), which are highly relevant to the low-energy electronic properties. The energy gaps and EMs at the high-symmetry $K$ points are listed in Table 2. Our result shows that the indirect (direct) energy gap obtained from the TB model and quasi-particle energy calculations are 2.79 eV (3.30 eV) and 2.75 eV (3.38 eV), respectively. We also find that both electrons and holes can be well described by the TB model. The EM of electrons at the CBM is 0.41 $m_0$, close to the value of 0.39 $m_0$ from $GW_0$ calculation. As the 1$^{st}$ and 2$^{nd}$ valence bands are degenerated at the Γ point, we show the EMs of holes for both bands in the table where one can find that the values are calculated to be 0.44 $m_0$ and 0.06 $m_0$, respectively, which are in good agreement with the corresponding $GW_0$ results of 0.40 $m_0$ and 0.08 $m_0$.

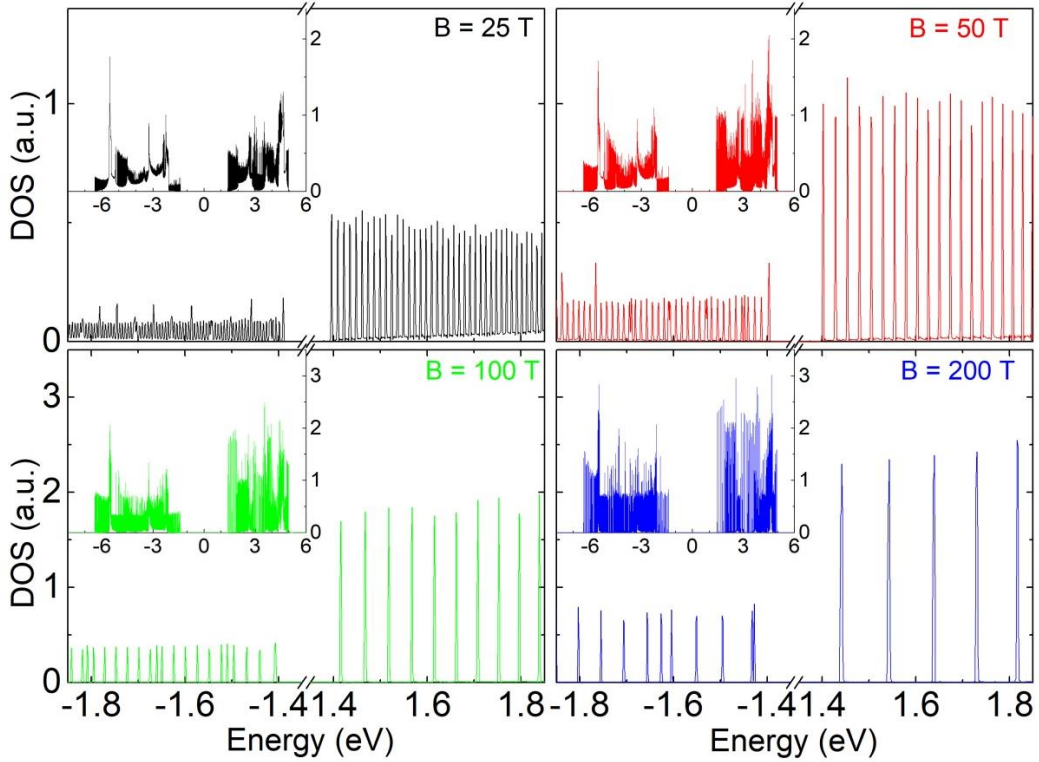

**Figure 4** DOS of ML-As under perpendicular magnetic field. The DOS is exaggerated from the inset with less denser peaks. Black, red, green and blue presents the magnetic field strength of 25 T, 50 T, 100 T and 200 T, respectively.

Now, we extend our model to the case of an external magnetic field through Peierls substitution in which the hopping term $t_{ij}$ in Eq. (1) is replaced by $t_{ij}e^{-i\frac{2\pi}{\Phi_0}e\int_{R_i}^{R_j}\mathbf{A}\cdot d\mathbf{l}}$, where $\Phi_0 = hc/e$ and $\mathbf{A} = (-By, 0, 0)$ are the flux quantum and the vector potential in the Landau gauge, respectively. Then, the DOS is calculated for a large sample containing 2*1000*2000 sublattices by means of the tight-binding propagation method, [30] where explicit evolution of the current operator is considered instead of diagonalization of large matrices. When the magnetic field **B** is applied perpendicular to the sheet, the energy levels become quantized, leading to discrete DOS as the Landau Level (LL) peaks. Because the electrons and holes are not symmetric in ML-As, the LLs for them exhibits different distance as shown in Figure 4. It is also noted in the low-energy region that there are two sets of LLs for holes, and we contribute this to the 1st and 2nd valence bands which are degenerated at the Γ point. Generally, these LL spectrum can be simply described by a linear function of the Landau index $n$ within an effective $\mathbf{k}\cdot\mathbf{p}$ model as

$$\epsilon_{n,s}^{kp} = \epsilon_s + \frac{seB\hbar}{m_0}\left(n+\frac{1}{2}\right)\omega_s, \quad (10)$$

where $s = \pm 1$ denotes the conduction and valence bands, $\epsilon_{+/-} = \epsilon_{c/v}$ is the energy at the conduction and valence edge, and $\omega_s = m_0/m_s^*$ is the relative ratio between free electron and the real system. However, our result shows that the carrier EMs in ML-As is energy-dependent [see Figure 3b], indicating a variable $\omega_s$ in Eq. (10), which in return is invalid for the electronic properties investigation with extremely-high magnetic field. The LLs spectrum with magnetic field up to 50 T obtained from our TB calculations is shown in Figure 5. In a feasible magnetic field strength of B = 25 T, one can see that the LLs spectrum follows the linear dependence on the magnetic field and level index which is fitted with $\epsilon_{+/-} = $ 1.382 eV (-1.383 eV) and $\omega_{+/-} = 2.104$ (2.256), respectively.

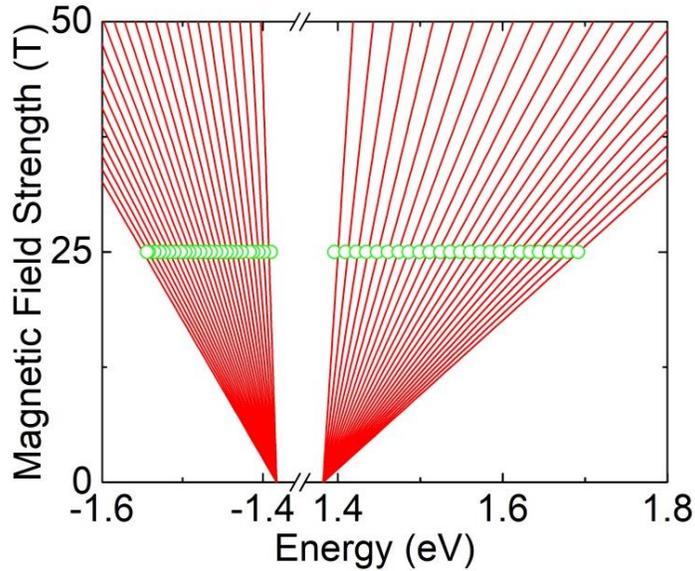

**Figure 5** LL spectrum of ML-As in high magnetic field. The lowest fifty LLs calculated from the linear fitting of Eq. (10) are indicated by the red dashed lines. The green circles are LLs obtained from tight-binding propagation method.

Then, we further investigate the optical properties of ML-As by calculating the frequency-dependent optical conductivity $\sigma_{\alpha\beta}(\omega)$ using the Kubo formula as implanted in our tight-binding propagation method. Here, we choose the same sample as in the LLs calculation with ~$10^6$ atoms and focus on the diagonal components of $\sigma_{\alpha\beta}(\omega)$ only. Our result of the optical conductivity $\sigma_{xx}$ is shown in Figure 6 with the magnetic field strength varies from 0 T to 100 T. When no external magnetic field is applied (**B** = 0 T), a sharp increase of $\sigma_{xx}$ appears around $\omega/t = 3.25$ eV which corresponds to the direct optical transition from the 1st VB to the 1st CB at the Γ point. Thus, our TB model is further validated by giving a reasonable optical gap comparable to the quasi-particle gap (3.38 eV) obtained from $GW_0$ calculation within random phase approximation.

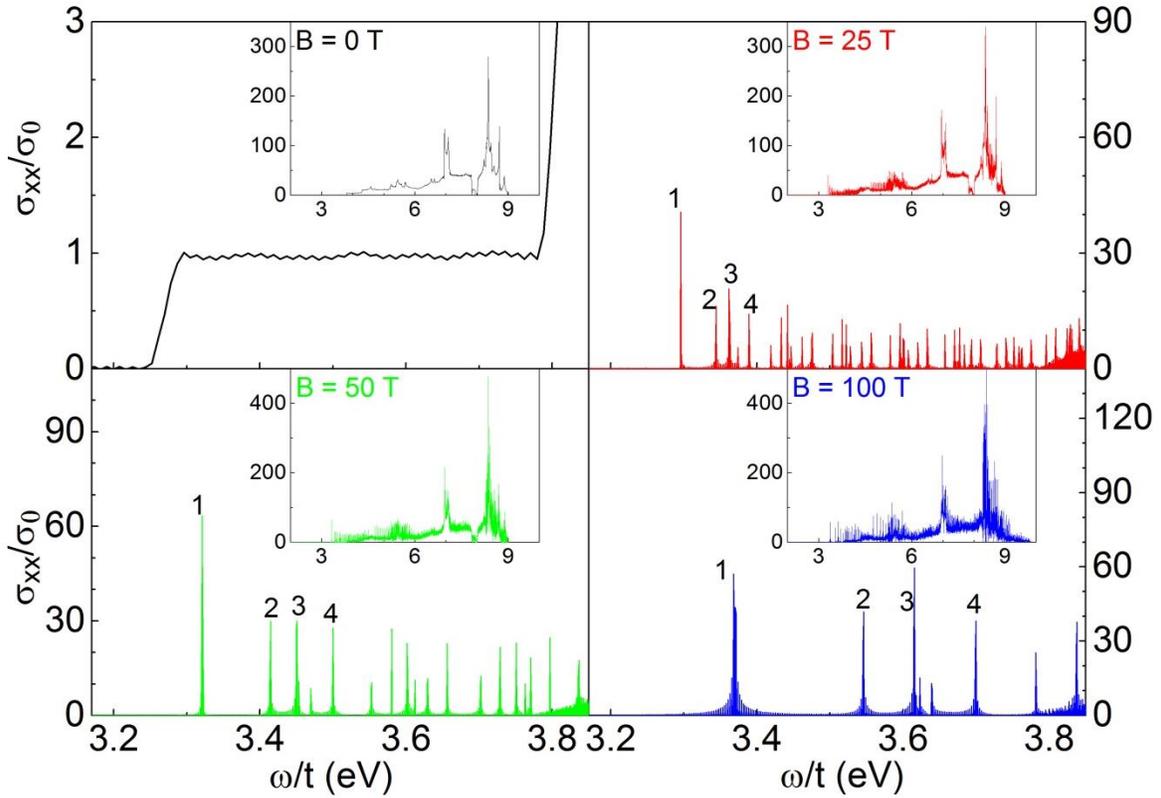

**Figure 6** Magneto-optical properties of ML-As. The optical conductivity spectrum of ML-As is calculated with **B** = 0 T (black), 25 T (red), 50 T (green) and 100 T (blue), respectively. The inset shows the optical conductivity in a much wider energy window. $\sigma_0 = e^2/4\hbar$ is the universal optical conductivity.

In the presence of a perpendicular magnetic field, the continuous optical conductivity in the low-energy region becomes quantized with discrete values. To get a clear image of the optical transition, we list the first few peaks of the optical conductivity in Table 3. As the strength of the magnetic field increases, the $n$-th optical conductivity peak shows a blue shift accompanying by a gap broadening. Taking the first four peaks in Figure 6 as an example, when the magnetic field varies from 25 T to 100 T, the 1$^{st}$ and 2$^{nd}$ peak of the optical conductivity increases from 3.296 eV and 3.344 eV to 3.368 eV and 3.546 eV, respectively. Our analysis on the optical transition also shows that not all the transition is permitted between electrons and holes in ML-As. Here, we show that the 1$^{st}$ peak with **B** = 50 T originates from the possible transition of carriers in VB$_{34}$-CB$_7$ and VB$_{15}$-CB$_{16}$, where VB$_i$ and CB$_j$ is the $i$ th and $j$ th LLs of the VB and CB, respectively. To this end, we conclude that the optical conductivity of ML-As with large size can be tuned effectively by the external magnetic field.

**Table 3** Optical gap of ML-As under magnetic field. The first four peaks of the optical conductivity calculated from our TB model are listed with **B** = 25 T, 50 T and 100 T, respectively.

| B (T) | Optical Gap (eV) | | | |
|---|---|---|---|---|
| | 1st | 2nd | 3rd | 4th |
| 25 | 3.296 | 3.344 | 3.362 | 3.389 |
| 50 | 3.321 | 3.415 | 3.451 | 3.500 |
| 100 | 3.368 | 3.546 | 3.618 | 3.700 |

**Conclusion**

In summary, we have proposed an effective Hamiltonian for monolayer arsenene which is derived from six $p$-like orbitals based on the partially self-consistent GW$_0$ approach. Using this tight-binding model, we can reproduce the electronic and optical properties of ML-As obtained from first-principles calculations, especially, in the low-energy region. By fitting the numerical results from our tight-binding propagation method, we find in ML-As a linear dependence of Landau levels on the perpendicularly applied magnetic field **B** and level index $n$. The optical conductivity spectrum of ML-As also shows a blue shift of the optical conductivity peaks accompanying by a broadening of the gap when the strength of magnetic field increases.

## Author information

Correspondence should be addressed to j.yu@science.ru.nl and s.yuan@whu.edu.cn .

## Acknowledgements

Yu acknowledges financial support from MOST 2017YFA0303404, NSAF U1530401 and computational resources from the Beijing Computational Science Research Center. Katsnelson acknowledges financial support from the European Research Council Advanced Grant program (Contract No. 338957). Yuan acknowledges financial support from Thousand Young Talent Plan (China).